\documentclass[preprint,double-spaced,showpacs,amsmath,amssymb]{revtex4}\textwidth=14.5cm \textheight=24cm
\usepackage{amssymb,graphicx}
\usepackage{dcolumn}

\def\P3{{\cal P}_t}
\def\J3{{\cal J}}
\def\T3{{\cal T}}

\def\beq{\begin{equation}}
\def\eeq{\end{equation}}
\def\bar{\begin{array}[b]}
\def\barc{\begin{array}}
\def\bart{\begin{array}[t]}
\def\ear{\end{array}}

 1  1
 1 
scaled\magstephalf 
 
scaled\magstephalf

\begin{document}
\thispagestyle{empty} \vspace*{0.5 cm}
\begin{center}
{\bf Single-angle to multi-angle transition in the collective flavor dynamics of neutrinos in supernovae.}
\\
\vspace*{1cm} {M. Baldo and V. Palmisano}\\
\vspace*{.3cm}
{\it INFN, Sezione di Catania}\\
{\it Via S. Sofia 64, I-95123, Catania, Italy} \\
\vspace*{.6cm} \vspace*{1 cm}
\end{center}

\begin{center}{\bf ABSTRACT} \\\end{center}

In this paper we study in detail the dynamics of flavor transformation for neutrinos propagating in the very dense environment of astrophysical
compact objects as Type II supernova in post collapse phase and proto-neutron stars. The analysis is based on the formalism by Strack and Burrows,
who introduced the generalized Boltzmann equation for Wigner phase space density. In appropriate limits the formalism reduces to the usual evolution
equations for the wave functions or for density matrix elements of Liouville equation. We incorporate the most important aspects of neutrino
propagation physics: the phenomenology of standard oscillations with MSW resonance induced by ordinary matter, collective behavior due to
self-interaction, which can produce bipolar and synchronized flavor oscillations, whose relevance is recognized by recent literature, the combination
of these effects with collisions including scattering, emission and absorption of neutrinos. In our numerical simulations a framework with only two
flavors are adopted together a geometry with spherical and azimuthal symmetry. An expansion into Legendre polynomials is adopted. It has the
advantage on one hand of incorporating the angular dependence, and on the other hand of simplifying the treatment of the self-interactions. This
method turns out to be sufficiently stable for the considered applications and numerically equivalent to other developments in term of multipoles
series, as discussed in the literature.

\par

We focus our analysis on the transition from single-angle to multi-angle behavior of the flavor content in the neutrino outflow. We include both
baryon matter and electrons and follow the change of the flavor dynamics as their density is varied. We distinguish regions close to the neutrino
sphere and more distant regions and in each region we estimate the critical matter density above which the neutrino flavor dynamics is suppressed. We
argue that the region where the transition can occur will move towards the neutrino sphere and the neutrino signal will be enhanced in the later
stage of the emission epoch.

\vskip 0.3 cm


\section{Introduction}

Today it is commonly accepted that the study of neutrinos from supernovae with core collapse is an important contribution to development of knowledge
about elementary particle physics and also to the understanding of the evolution mechanism of astrophysical sources
\cite{beth,taka,thom,bur1,full,sigl}. In fact the supernova in the post bounce stage are substantially similar to a black body emitting neutrinos of
all three flavors, whose propagation in the very dense stellar material is highly sensitive to parameters such as mass hierarchy or the mixing
angles. It may be useful in this sense to distinguish two different types of phenomena related to the neutrino flavor conversion in terms of distance
from the source and discuss specifically the effects of conversion to long distance and short distance. The first conversion phenomena are caused by
nuclear weak interactions of neutrinos with the constituents of ordinary matter such as electrons, protons and neutrons via charged and neutral
currents, even if the typical average value for neutrinos energy of about 10 MeV is much less than the mass of the $\mu$ and $\tau$ leptons and only
the interaction by neutral current takes place. These interactions can be decisive in affecting the conversion of the flavor and occur at a a
distance from the core $r \gg 10^3$ km \cite{bil1,bil2,bald,bal1,fog1,fog2}. The main phenomenon is the celebrated MSW resonance, that was decisive
to solve the neutrino puzzle in the solar case \cite{wolf,smir}. The second conversion phenomenon instead occurs at much shorter distances from the
core of star, $r \sim$ $10 - 150$ km, where concentration of produced neutrinos is so high to make self-interactions not negligible. Numerical
simulations in recent studies show that in this situation interesting effects emerge, all characterized by the collective behavior of neutrinos and
antineutrinos which are coupled regardless of their energy or propagation direction \cite{bal2,fog3,kos2,raf1,raf2,raf3,raf4,qian}. These effects are
completely different from ordinary oscillations in matter because they involve colliding neutrinos with a cross section that depends on the angle
between their momenta. These particular features introduce a set of coupled non-linear equations and makes prohibitive an exact solution of the time
dependent kinetic equations in the general case \cite{raf5,raf6}. However reliable approximations have been developed, which have lead to a detailed
studies of the neutrino flavor dynamics.

\par

One of the main results \cite{raf1,raf2,raf3,raf4,qian} of these studies is the prediction that at a certain distance from the neutrino sphere a
transition from coherent flavor oscillations to incoherent ones should occur. The coherent oscillations correspond to flavour oscillations that are
in phase among the different direction along which neutrino are propagating, while in the incoherent ones the different directions are decoupled and
the flavour oscillations are out of phase. Another predicted process is a sharp transition from the synchronized oscillations to the so called
``bipolar'' ones, that should mainly correspond to the mutual conversion of neutrino-antineutrino pairs of one flavour to another one. The two
transitions can appear to occur simultaneously, and they are the subject of intense theoretical studies. All these phenomena can occur at not too
large distance from neutrino sphere, since otherwise the neutrino density becomes too small, due to the spherical geometry, and the momenta of the
interacting neutrinos are necessarily focused along the radial direction, which reduces the effective neutrino-neutrino interaction strength. These
processes are strongly affected by the matter physical conditions, besides the neutrino and antineutrino flavor compositions. In particular the
presence of electrons can reduce the effective mixing angle in the matter to a very small value. For high enough electron density it is expected that
all flavor oscillation processes will be blocked by the increase of the electron effective potential on the neutrinos due to neutral weak current
interaction. Furthermore, closer to the neutrino sphere neutrino scattering and absorption can be relevant and modify the overall physical picture.
Therefore in this case they should be included in the theoretical simulations. At larger distance the matter density decrease rapidly and both
scattering and absorption become negligible, while electrons can still play a role.

\par

In this paper we analyze the behavior of neutrino flavor content by Boltzmann-like equations, following the formulation of Strack and Burrows
\cite{stra}. We distinguish between the region close to the neutrino sphere, where the role of matter can be of decisive relevance, and the region
away from the neutrino sphere, where the neutrinos flow essentially freely and along straight line trajectories, where both neutrino and electron
densities are decreasing. We examine the physical conditions at different distances from the neutrino sphere and simulate the flavor evolution under
realistic assumptions within the intrinsic uncertainties of the problem.

\section{The formalism}

\par

Different research groups have implemented numerical schemes capable of solving the complex equations governing the dynamics of flavor in a general
framework, using different techniques. In particular multi-angle simulations have been used, where each trajectory of neutrinos is followed. It has
been also explored the possibility of reducing the complexity of the calculation using appropriate developments of the angular dependence in a series
of polynomials, in such a way that the final equations are simpler to solve \cite{raf4}. This is particularly useful when the initial angular
distribution of the neutrinos is highly non-isotropic. These studies did not discuss how the phenomenon of flavor modulation due to the above
mentioned processes is influenced by the effects of neutrino reactions in the background matter, i.e.  neutrino scattering or emission and
absorption. In this paper we include in the numerical simulations also the effects of dynamical decoherence caused by collisions of the neutrinos
with the ordinary matter's background. The main processes correspond to protons and neutrons that scatter, absorb and emit particles in addition to
electrons that can drastically alter the effective flavor mixing angle. The evolution equations that we use in this work are the generalized
Boltzmann equations of Ref.~\cite{stra} that can simulate the complex oscillation phenomenology of neutrinos and simultaneously take account the
neutrino reactions in matter. We solve these equations using the expansion in Legendre polynomials and we obtain a suitably truncated set of
equations that can be numerically handled by the standard software that use an adaptive Runge-Kutta methods. Our main goal is, on one hand, to study
neutrino flavor evolution near the neutrino sphere, where the presence of collisions and absorption, as well as the electron component, could
generate decoherence and tends to suppress the flavor conversion and oscillations. On the other hand we extend the study to larger distance from the
neutrino sphere, where neutrinos flow freely but the electrons can still suppress the flavor oscillations.

\par

In the Appendix details are given on the semi-classical Boltzmann-like equations of Ref.~\cite{stra} and their expansion in Legendre polynomials.
Here we sketch the main elements of the theory and its numerical implementation. The basic quantity is the Wigner-Ville distribution in phase space.
It is a quasi-probability distribution introduced to study quantum corrections to classical statistical mechanics which links the wave function, that
appears in Schr\"odinger's equation to a probability distribution in phase space. In second quantization formalism it can be defined as

\begin{equation}\label{wigner-density}
\rho \left( {{\bf{r}},{\bf{p}},t} \right) = \int {\frac{{d^3
{\bf{r'}}}}{{\left( {2\pi \hbar } \right)^3 }}e^{ -i{\bf{pr'}}} \psi
^\dag  \left( {{\bf{r}} - \frac{1}{2}{\bf{r'}},t} \right)} \psi
\left( {{\bf{r}} + \frac{1}{2}{\bf{r'}},t} \right),
\end{equation}

\noindent where $\psi ^\dag$ and $\psi$ are creation and annihilation operators.

Recently \cite{stra} a formalism was developed to describe transport processes in terms of Wigner functions, which is able to incorporate the quantum
phenomenon of flavor conversion of neutrinos. The basic equations of the theory can be derived heuristically merging Boltzmann equation with
Heisenberg equation. They describe the spatial-time evolution of the matrix elements of the Wigner density. Taking the matrix elements in Fock space,
considering for simplicity only two neutrino flavors, one can write according to Ref.~\cite{stra}

\begin{equation}\label{F-matrix}
{\cal F} = \left\langle {n_i } \right|\rho \left| {n_j }
\right\rangle  = \left( {\begin{array}{*{20}c}
   {f_{\nu _e } } & {f_{e\mu } }  \\
   {f_{e\mu }^ *  } & {f_{\nu _\mu  } }  \\
\end{array}} \right),
\end{equation}

The Boltzmann-like equation can be schematically written

\begin{equation}\label{eq:Boltz}
\frac{\partial {\cal F}}{{\partial t}} + {\bf{v}}\cdot\frac{\partial {\cal F}}{\partial {\bf{r}}} +  {\bf{\dot p}}\cdot\frac{\partial {\cal
F}}{\partial {\bf{p}}} \,=\,  I\{{\cal F}\}
\end{equation}

\noindent where the left hand side describes the free streaming of the neutrinos and the right hand side includes the processes that involve
neutrinos, i.e. the flavor mixing matrix, the neutrino-neutrino interaction as well as scattering, absorption and emission by matter. The time
derivative of the momentum $\bf{p}$ is caused by the gravitational field and can be neglected. Despite this simplification, the complete solution of
these equations is extremely difficult. Two main approximations can be used to obtain from these equations the neutrino flavor evolution along their
path away from the star. One can assume that the neutrino emission from the star is quasi-stationary. This is a good approximation if the processes
that are described by the simulations have a time scale much shorter than the total neutrino emission time. In this case one can neglect the time
derivative in Eq.~(\ref{eq:Boltz}) and the neutrino flavor evolution is followed along the radial coordinate, assuming spherical symmetry. In
particular \cite{ful1,ful2,ful3,ful4,ful5,ful6,este} one can assume that scattering, absorption and emission are negligible and that neutrinos flow
freely along straight line trajectories. Along each one of them the flavor content evolves due to the neutrino-neutrino interaction and the effective
mixing angle in matter.

\par Another possibility is to approximate locally the time evolution of the distribution functions, considering the local matter as homogeneous and
in planar geometry. This is justified if, within the characteristic time of the considered processes, neutrinos can travel for a distance shorter
than the length scale of the star matter and neutrino density profiles. In this case one considers the time derivative and neglect the coordinate
derivative of the distribution functions, and the equations are solved locally at a given distance from the neutrino sphere. For simplicity we will
adopt this second scheme, checking that the underlying assumptions are indeed well satisfied. However we will implement the method by including the
increasing focusing of the neutrino flow at a a given point of the radial direction as the distance from the neutrino sphere increases. The solution
of the equations based on the quasi-stationary assumption within the same numerical method of Legendre polynomial expansion is left to a future work.

\par

If neutrino scattering is negligible, the focusing of the neutrino flow is dictated by simple geometrical considerations \cite{duan}. If $R$ is the
neutrino sphere radius, at a given point at a radial distance $r > R$ the flow of the emitted neutrino is restricted within an angle $\theta_{max}$
with respect to the radial direction. According to Ref.~\cite{duan}, one has

\beq \cos\theta_{max} \,=\, \sqrt{1 \,-\, \frac{R^2}{r^2}} \label{eq:tmax}\eeq \noindent and each angular direction $\theta_p \,\leq\, \theta_{max}$
is related to the angle of emission $\theta_{0}$ from the neutrino sphere

\beq \mu_p \,=\, \sqrt{1 \,-\, \frac{R^2}{r^2} ( 1 \,-\, \mu_0^2 ) } \label{t0}\eeq

\noindent where $\mu_p \,=\, \cos\theta_p$ and $\mu_{0} \,=\, \cos\theta_{0}$. The angular distribution of neutrinos must be restricted within this
angular width at a given radial distance $r$, see figure 1 in Ref.~\cite{duan}.

\section{Flavor dynamics close to the neutrino sphere}

\begin{figure}[t]
\begin{center}
\includegraphics[width=0.95\textwidth]{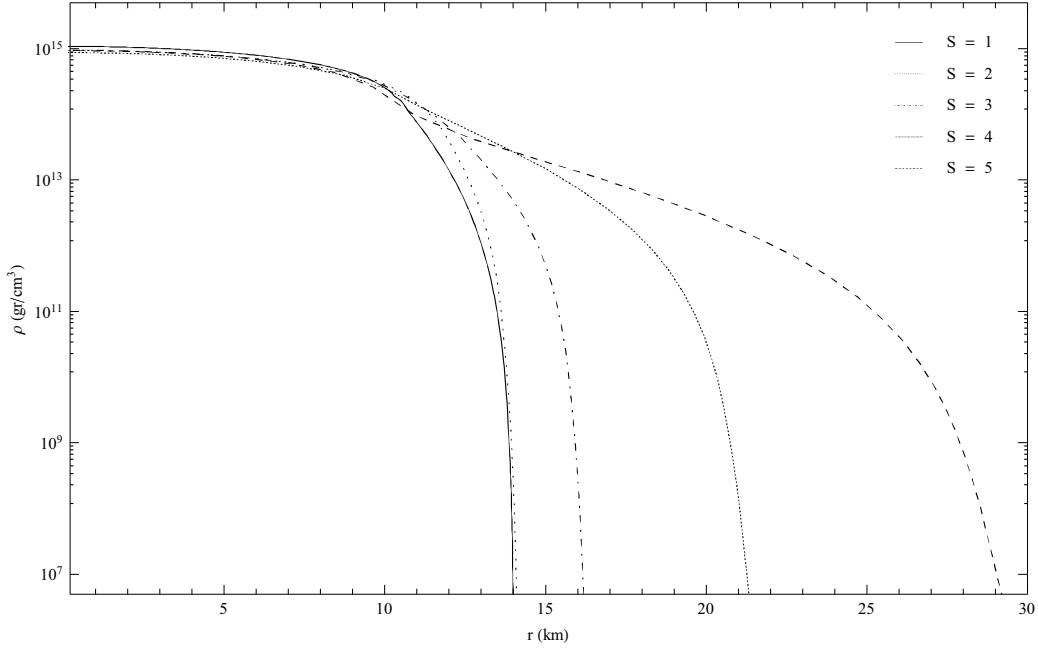}\caption{Baryon density profiles of a proton-neutron star of mass
$M/M_\odot = 1.4$ for different entropy. The core entropy is fixed at S = 1 while the envelope entropy  is taken, from the lower to the higher
profiles, as S = 1, 2, 3, 4, 5.} \label{fig:fig1}
\end{center}
\end{figure}

We start our analysis considering the region near the neutrino sphere, where the matter density is high enough to produce relevant effects on the
evolution of neutrino flavor and the neutrino density is the highest one. This region is not only relevant by itself but also it can affect the
initial conditions for the subsequent neutrino propagation. Since the size of the neutrino sphere is different for different flavors, during the
emission process there are regions with an imbalance of neutrino contents with respect to the thermodynamic equilibrium. It is then of interest to
study the flavor evolution under these physical conditions and to estimate the characteristic time scale of the possible flavor conversion. As
physical parameters we consider the total neutrino density $\rho_\nu \,=\, 10^{34}$ cm$^{-3}$ and vary the matter density with typical values in the
interval from $10^6$ to $10^{13}$ $g/$cm$^{3}$. The electron fraction, if included, is kept to the value $0.4$. This is justified by microscopic
quasi-static calculations of the structure of the after-bounce proto-neutron stars. For future discussion Fig.~\ref{fig:fig1} reports a sample of
these matter density profiles at different entropy of the envelope calculated following the scheme of Ref.~\cite{burg}. The calculation assume
neutrino trapping, thermodynamic and chemical local equilibrium, and an entropy per particle $S = 1$ in the core and $S = 1, 2, 3, 4$ in the
envelope. The lowest entropy profiles should correspond to the latest stage of the neutrino emission period. The results are in line with dynamical
simulations of the Livermore group, see Fig. 1 of Ref.~\cite{schi}. Despite the quasi-static calculations do not extend to the more far tails of the
profiles and the two sets of calculations use two  different types of scheme, they show a very similar trend  and their overall evolution suggests
the range of density values that the profile should cover during the neutrino emission epoch.

\par

\begin{figure}[t]
\begin{center}
\includegraphics[width=0.85\textwidth]{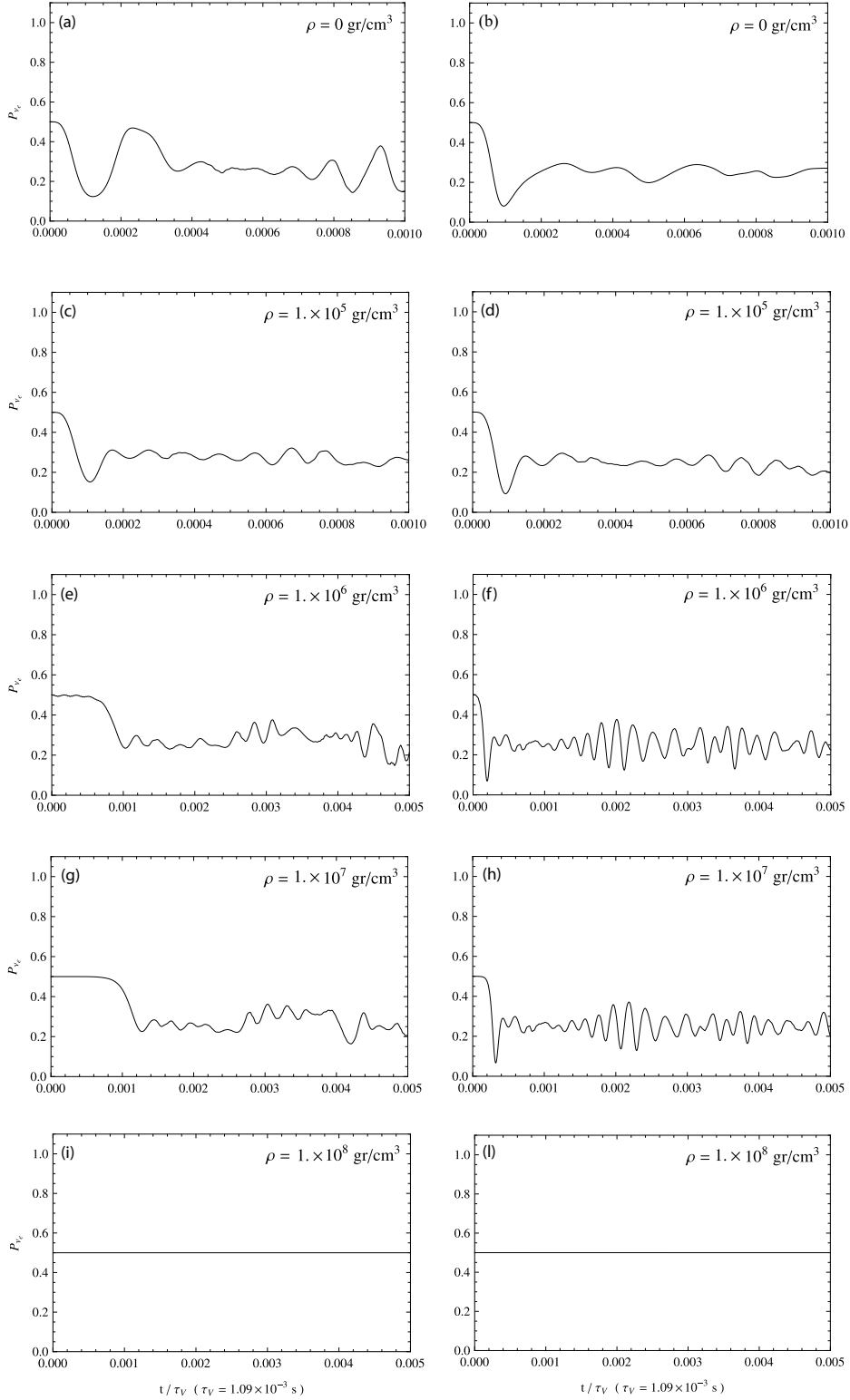}
\caption{Time evolution of $\nu_e$ fraction at radial distance of $15$ km and different densities matter values. Energy is fixed to $10$ MeV and
neutrino density to $10^{34}$ cm$^{-3}$. Time is in unit of vacuum oscillation period $\tau_V$. Left: normal hierarchy. Right: inverted hierarchy. }
\label{fig:fig2}
\end{center}
\end{figure}

\begin{figure}[t]
\begin{center}
\includegraphics[width=0.9\textwidth]{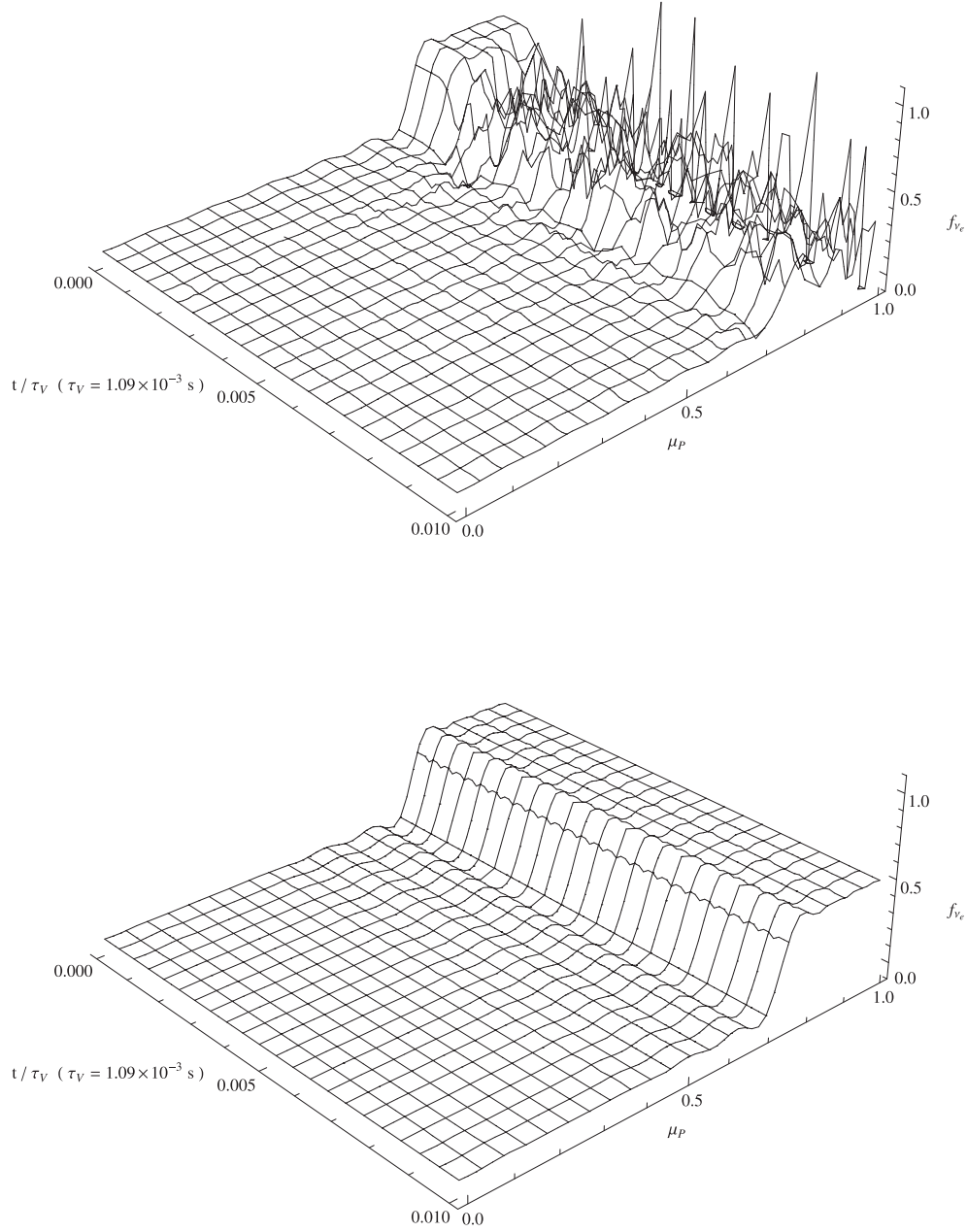}
\caption{Angular distribution of $\nu_e$ content in function of time with same parameter of Fig.~\ref{fig:fig2} in normal hierarchy. Quantity
$f_{\nu_e}$ is normalized to total neutrino density and $\mu_p=\cos(\theta_p)$, with $\theta_p$ the angle with respect to the radial direction Upper
plot: density matter fixed to $10^7$ gr/cm$^3$. Lower plot: density matter fixed to $10^8$ gr/cm$^3$. } \label{fig:fig3}
\end{center}
\end{figure}

For simplicity we consider a two-flavor framework, with parameters corresponding approximately to the electron and atmospheric neutrinos, see
Table~\ref{tb:tab}, conventionally indicated as electron and muon neutrinos \cite{schw}. For absorption and emission processes one needs to specify
the neutrino temperatures. We use the indicative values reported in Table~\ref{tb:tab}, taken from Ref.~\cite{ful1}. First we consider equal number
of neutrinos and anti-neutrinos, but with an excess of electron flavor. Neutrinos are considered mono-energetic, and the equations are
correspondingly adapted as detailed in the Appendix. Following the scheme described in the previous section, in Fig.~\ref{fig:fig2} is reported the
time evolution of the fractional content in electron flavor of the neutrinos, with energy $E = 10$ MeV. This content ${P_{{\nu _e}}}$ is normalized
to the total number of neutrinos and anti-neutrinos. The angular distribution at the neutrino sphere is taken ``half isotropic'', i.e. uniform
angular distribution in the forward direction, and no backward emission, to simulate the flow from the neutrino sphere. We take the radius of
neutrino sphere equal to 10 Km and analyze the region around a radial distance of 15 Km. We start the simulation from an angular distribution
``focused'' according to Eq.~(\ref{eq:tmax}). The aim of the calculation is to check the stability of the angular distribution with time, in order to
estimate the time necessary for the appearance of the transition to the multi-angle regime. If the matter electron component is neglected as well as
the emission and absorption processes, the temporal evolution of $P_{{\nu _e}}$ is depicted in  Fig.~\ref{fig:fig2}a. The time unit is taken to be
the oscillation time period in vacuum $\tau_V$, as specified in the Appendix. In a time as short as $10^{-6}$ s there is a sharp drop. It corresponds
indeed to the transition from the ``single-angle'' evolution to the ``multi-angle'' one. This transition is apparent in the three-dimensional plot of
Fig.~\ref{fig:fig3}, where the angular distribution of the electron flavor content as a function of time is reported. Initially the angular
distribution stays stable with time. At the transition time, apparent in the upper panel of Fig.~\ref{fig:fig3}, the angular distribution changes
abruptly and becomes irregular and time dependent. It is to be noticed, however, that the angular distribution after the transition is not completely
random, but displays some structure. Furthermore it has to be kept in mind that the angular distribution is a probability density and it can acquire
arbitrary values. Notice also the short time necessary for the transition, which justifies the assumption of uniform matter that we are using. \par
If we introduce the electron component at increasing density, the signal disappears at some critical value. This is expected since the effective
mixing angle tends to vanishing small values as the electron density increases. Correspondingly, the angular distribution remains unchanged with
time, as depicted in the lower part of Fig.~\ref{fig:fig3}. The actual matter and electron density depends on the stage of the neutrino emission
epoch. At a fixed distance in the initial stage the matter density is likely to be larger than the critical one, while at a later stage it can be
smaller and the neutrino signal should then appears.

\par

In the second column of Fig.~\ref{fig:fig2} the same analysis is reported for the inverted hierarchy for the neutrino mass spectrum. A similar
behavior is observed, but the transition occurs at an early time.\par
\begin{figure}[t]
\begin{center}
\includegraphics[width=0.9\textwidth]{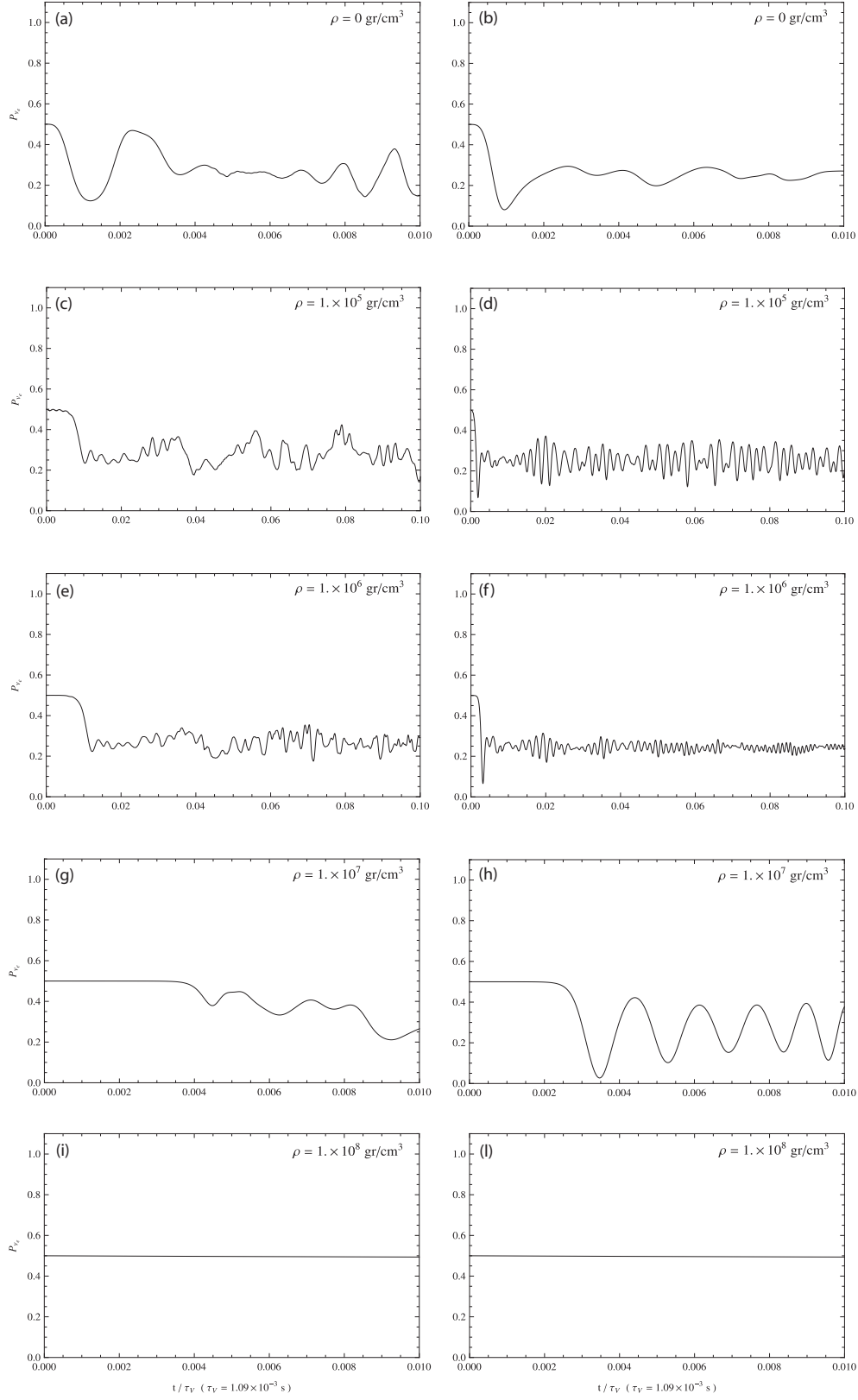}
\caption{The same as in Fig.~\ref{fig:fig2} but for neutrino density fixed to $10^{32}$ cm$^{-3}$.  }   \label{fig:fig4}
\end{center}
\end{figure}

\begin{figure}[t]
\begin{center}
\includegraphics[width=0.9\textwidth]{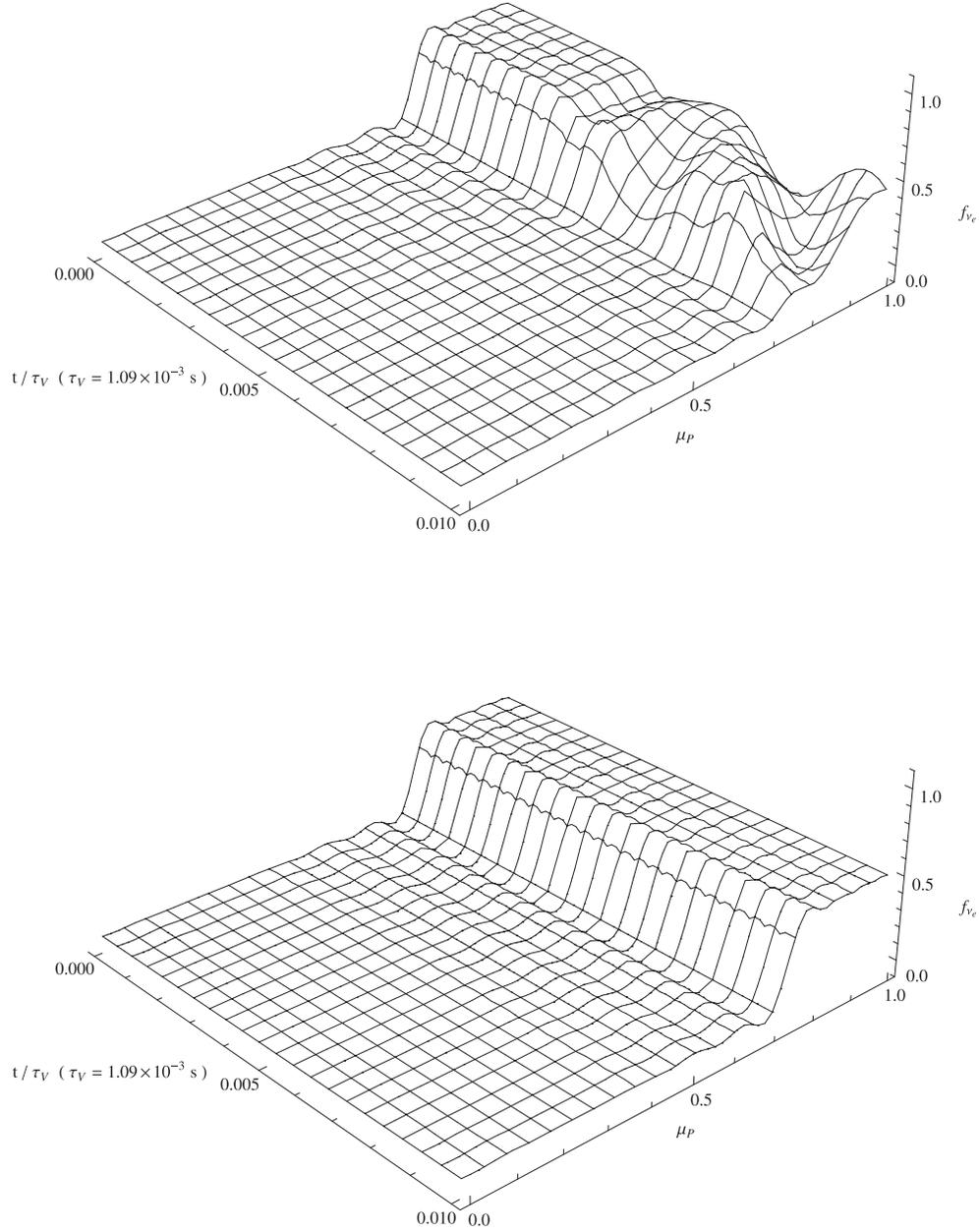}
\caption{Angular distribution of $\nu_e$ content as a function of time with the same parameters of Fig.~\ref{fig:fig4} in normal hierarchy. Quantity
$f_{\nu_e}$ is normalized to total neutrino density. Upper plot: density matter fixed to $10^7$ gr/cm$^3$. Lower plot: density matter fixed to $10^8$
gr/cm$^3$.} \label{fig:fig5}
\end{center}
\end{figure}

 The same type of calculations of Fig.~\ref{fig:fig2} are reported in
Fig.~\ref{fig:fig4} for a neutrino density $\rho_\nu \,=\, 10^{32}$ cm$^{-3}$. The observed trend is quite similar. However the transition is
obtained at a larger matter density and the time scale for the transition is longer. The time scale for the occurrence of the transition is still
acceptable for the validity of the method of analysis. One notices the scaling behavior : as the density decreases by two orders of magnitude, both
the transition density and the time for the transition increase by one order of magnitude. The corresponding angular distributions are reported in
Fig.~\ref{fig:fig5}. Also in this case the appearance of the transition is quite evident.

\par

\begin{figure}[t]
\begin{center}
\includegraphics[width=0.9\textwidth]{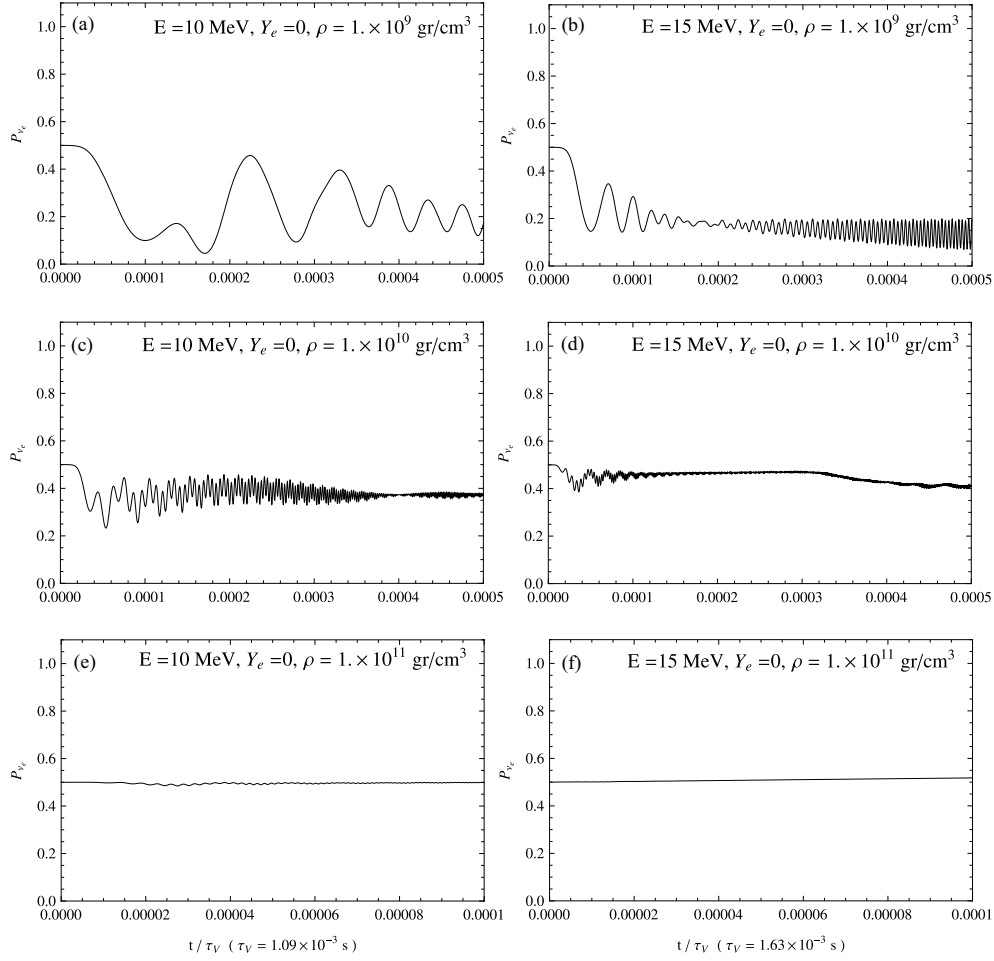}
\caption{The same as in Fig.~\ref{fig:fig2} but at higher matter density and for two different energies in direct hierarchy. Neutron fraction is
fixed to $1$ to put in evidence possible role of absorption, emission and scattering. Left: Energy is fixed to $10$ MeV. Right: Energy is fixed to
$15$ MeV. } \label{fig:fig6}
\end{center}
\end{figure}

\begin{figure}[t]
\begin{center}
\includegraphics[width=0.9\textwidth]{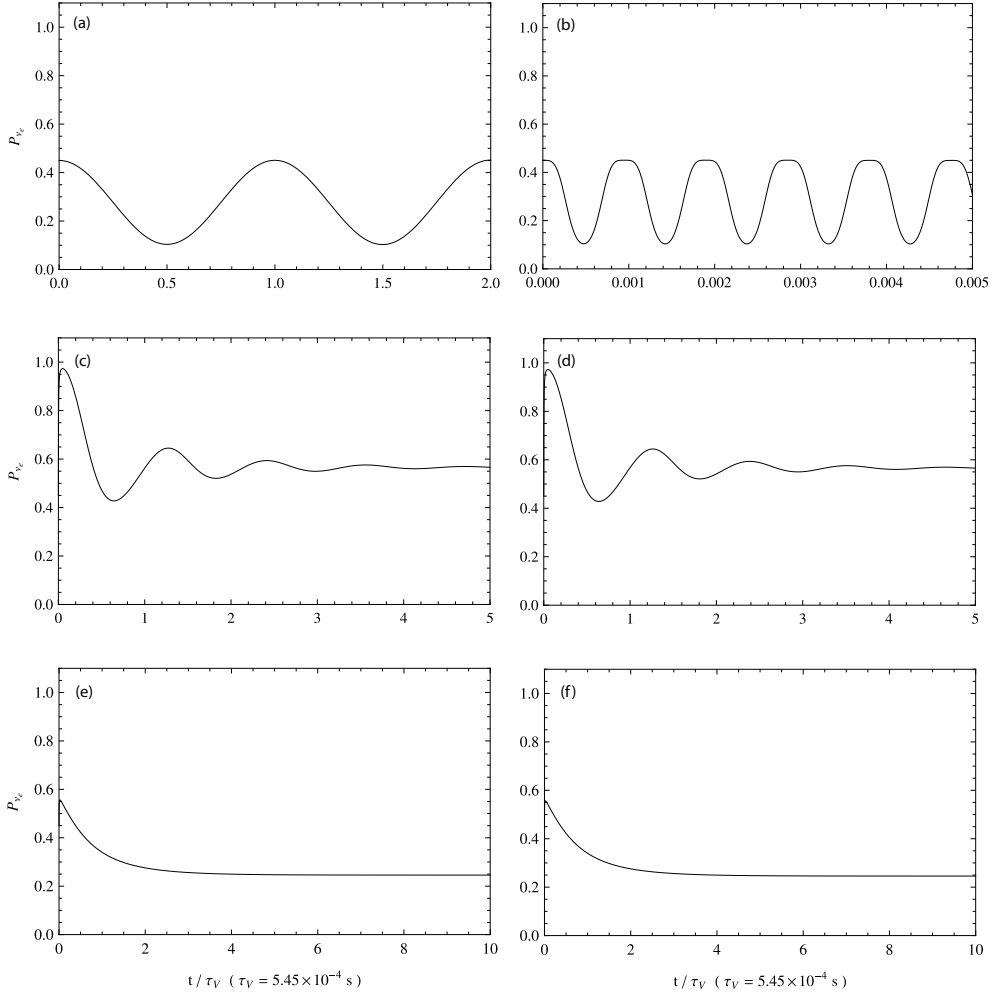}
\caption{Time evolution of $\nu_e$ fraction for initial isotropic angular distribution and different physical conditions. Energy is fixed to $5$ MeV,
neutrino density to $10^{32}$ cm$^{-3}$, density matter to $10^{11}$ gr/cm$^{3}$ and hierarchy is direct. Time is in $\tau_V$ unit. Panel (a): vacuum
oscillation, (b) self-interaction only, (c) only absorbtion/emission, (d) self-interaction and absorbtion/emission, (e,f) including also electrons
with or without self-interaction. } \label{fig:fig7}
\end{center}
\end{figure}

In all these cases we found that both absorption/emission and scattering processes have essentially no relevance. To put in evidence their possible
role, we suppressed the electron component and we repeated the calculations at the same neutrino density but at higher matter density and at two
different neutrino energies, assuming direct hierarchy. The results  can be seen in Fig.~\ref{fig:fig6}. They are able to suppress any flavor
dynamics, but only if the matter density is much higher. As a consequence the inclusion of the electron component, with the corresponding density,
would wash out any signal, as we have explicitly checked. Therefore the absorption/emission processes are expected to have very limited effects in
all cases. The scattering processes have a much higher time scale and play no role in any case. To illustrate further this point we consider the
evolution of the flavor content for an isotropic angular distribution, which would be more appropriate for the neutrinos inside the neutrino spheres
of both flavors. Accordingly, the matter density is taken at $10^{11}$ g/cm$^{3}$ and the neutrino density at $10^{34}$ cm$^{-3}$ . We assume the
same initial flavor contents as in the previous calculations, so that the simulations should give the characteristic time scale for the relaxation
toward equilibrium. For aim of comparison, we take the same temperature values for the two flavor neutrinos as in the previous analysis. As
reference, vacuum oscillations are reported in Fig.~\ref{fig:fig7}a. If only the neutrino self-interaction is included the usual bipolar oscillations
are observed, see Fig.~\ref{fig:fig7}b. If only absorption/emission are included, the time scale of the flavor dynamics increases by three order of
magnitude, see Fig.~\ref{fig:fig7}c. In this case one gets a relaxation toward equilibrium through damped oscillations. Despite the time scale is
drastically enhanced, it is still much shorter than the ones of typical macroscopic processes inside the neutrino sphere, as the trapping time or the
possible density oscillations. If in addition also the self-interaction is introduced, apparently no change is observed, i.e. the absorption/emission
processes dominate the flavor evolution, see Fig.~\ref{fig:fig7}d. Finally the inclusion of the electron component damps all the oscillations in the
evolution toward equilibrium with or without self-interaction, Figs.~\ref{fig:fig7}e, f. This analysis, even if only illustrative, indicates that the
time scale toward flavor equilibrium is much longer than typical bipolar oscillation period but shorter than the time scale of the macroscopic
processes inside the neutrino sphere.

We consider now the effect of the neutrino-antineutrino asymmetry. According to Ref.~\cite{este} if the flavor content is substantially different for
neutrino and antineutrino, the flavor dynamics and the transition from single-angle to multi-angle regime are suppressed. Assuming initially zero
flavor content of $\mu$ neutrino, i.e. ${{P_{{\nu _\mu }}}}={{P_{{{\overline \nu }_\mu }}}}=0$ at $t = 0$, the asymmetry can be specified by the
value of the parameter $\epsilon$ defined as

\begin{equation}
\epsilon  = \frac{{{P_{{\nu _e}}} - {P_{{{\overline \nu }_e}}}}}{{{P_{{{\overline \nu }_e}}}}},
\end{equation}

\noindent equal to the relative fraction of different neutrino and antineutrino electron flavor content.

\begin{figure}[t]
\begin{center}
\includegraphics[width=0.8\textwidth]{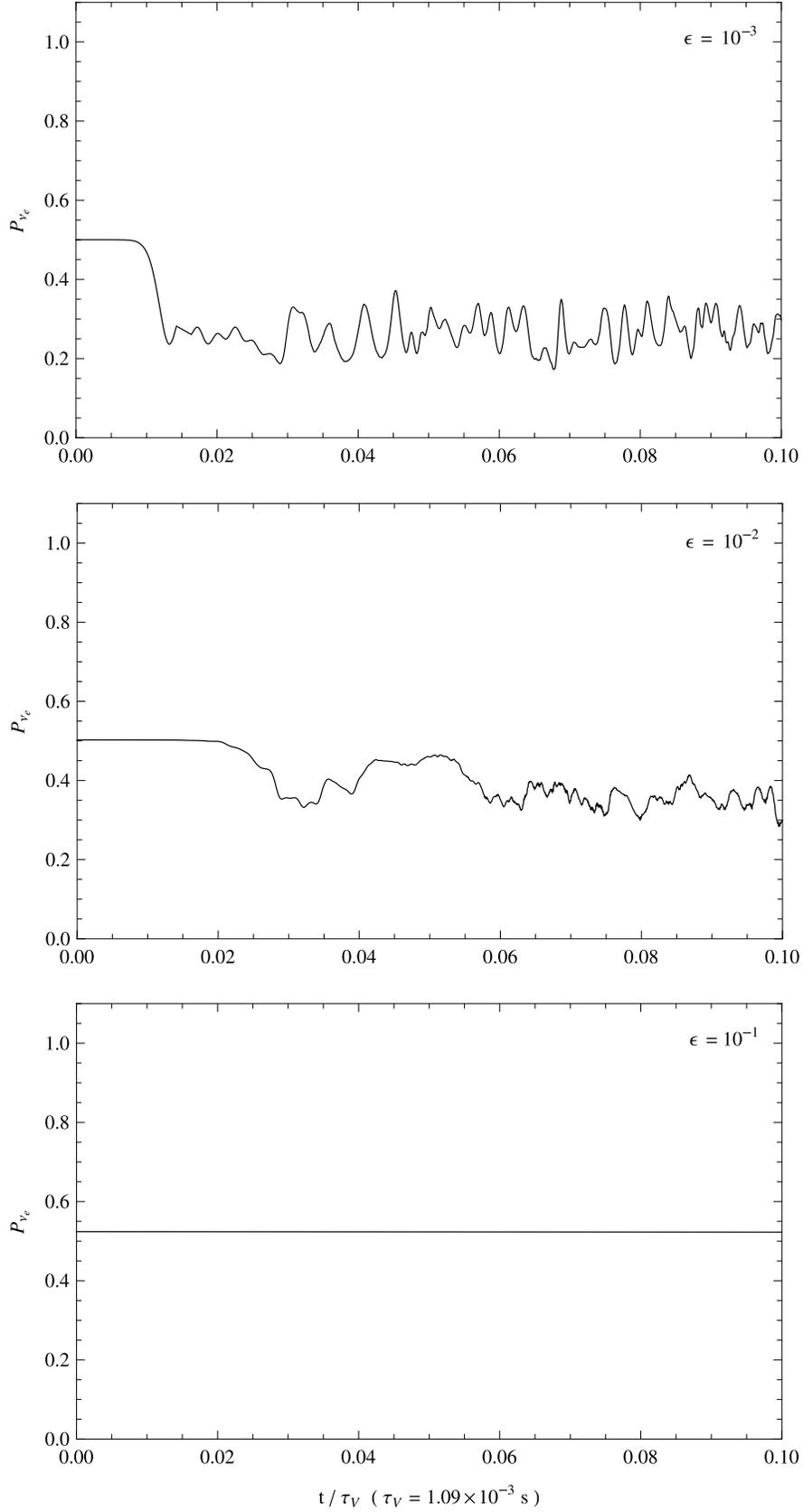}
\caption{Neutrino flavor dynamics at different values of the asymmetry parameter $\epsilon$, see the text for details.} \label{fig:fig8}
\end{center}
\end{figure}

We have investigated this effect under the physical conditions of Fig.~\ref{fig:fig7}e. The variation of the flavor time evolution at different
asymmetries is reported in Fig.~\ref{fig:fig8}. The suppression of decoherence seems to start around $\epsilon \,=\, 10^{-2}$. We therefore confirm
the phenomenon of decoherence suppression by asymmetry. The precise value of the critical value of $\epsilon$ depends of course on the effective
mixing angle and on the neutrino density. In general a value of the asymmetry as small as $10^{-1}$ or so is enough to fully suppress the flavor
coherence.

\section{Moving away from the neutrino sphere}
\begin{figure}[t]
\begin{center}
\includegraphics[width=0.9\textwidth]{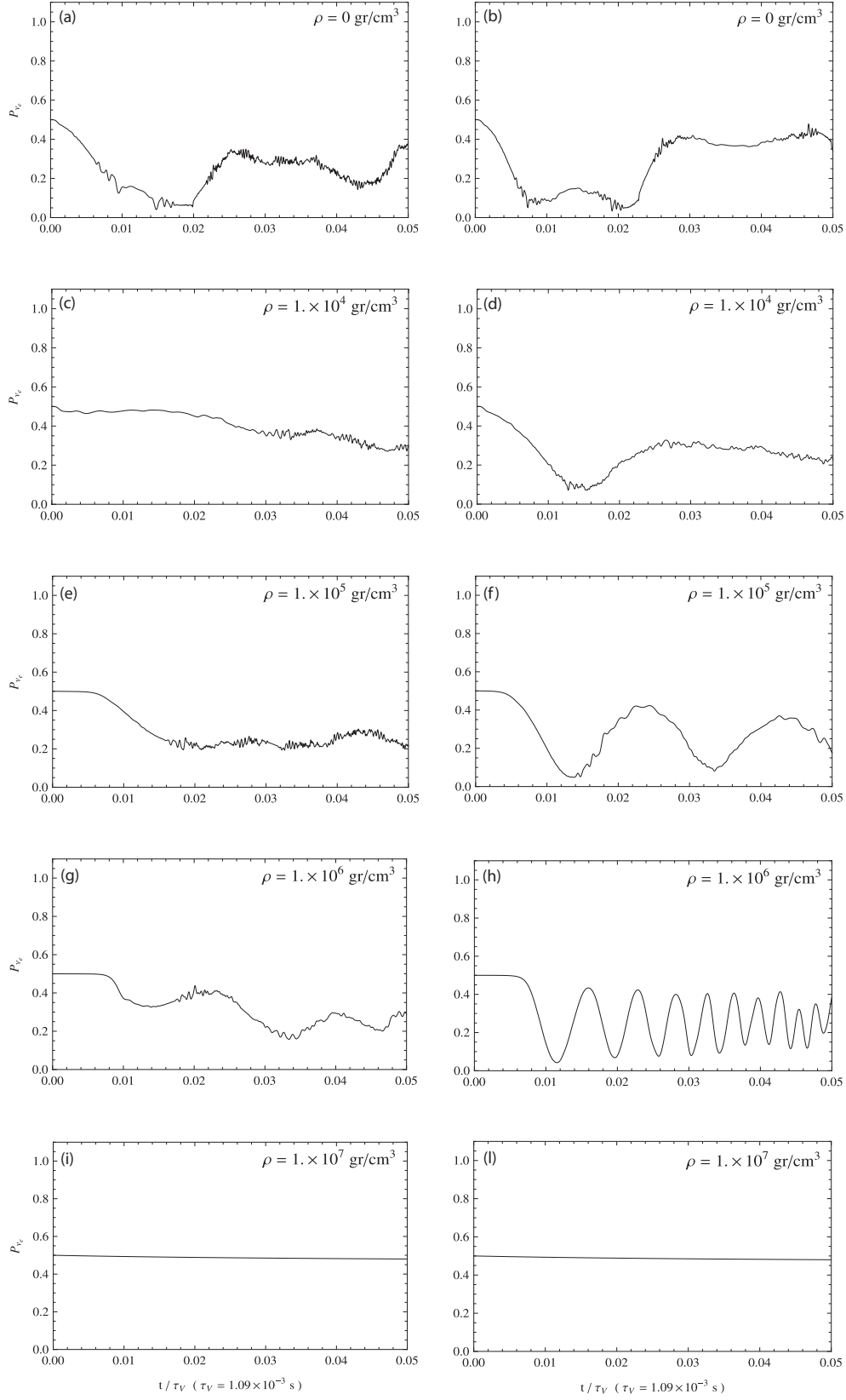}
\caption{The same as in Fig.~\ref{fig:fig4} but at radial distance of $50$ km. } \label{fig:fig9}
\end{center}
\end{figure}

\begin{figure}[t]
\begin{center}
\includegraphics[width=0.9\textwidth]{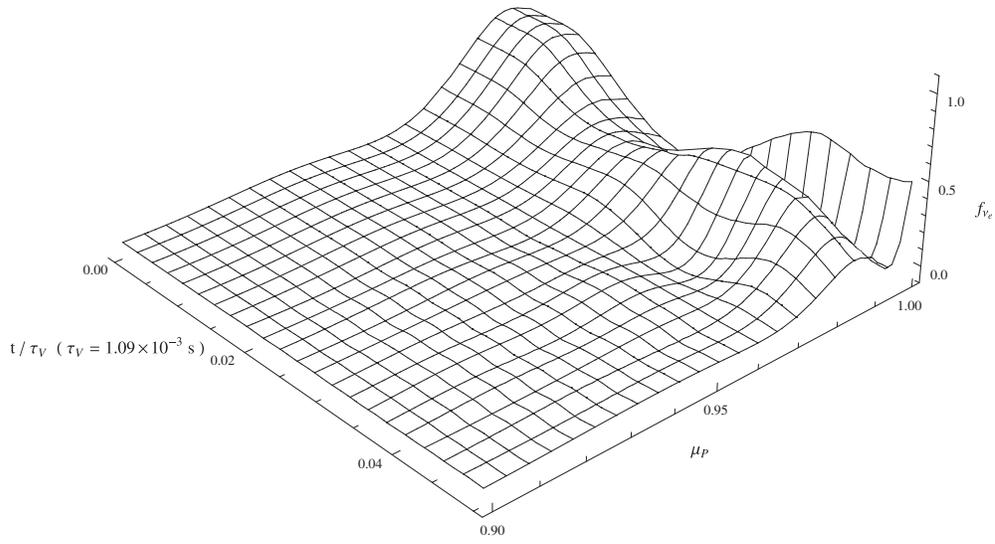}
\caption{Angular distribution of $\nu_e$ content as a function of time with the same parameters of Fig.~\ref{fig:fig9}e.} \label{fig:fig10}
\end{center}
\end{figure}

 We will consider in this section the flavor dynamics at larger distance from the neutrino sphere. With respect to the previous analysis the main
change that occurs is the degree of the focusing of the interacting neutrino beam. We do not report all the results, but we concentrate on the case
of neutrino density $10^{32}$ cm$^{-3}$. We consider the distance of $50$ km, away from the neutrino sphere. One can see from Fig.~\ref{fig:fig9}
that the time evolution at different matter density of ${P_{{\nu _e}}}$ is more irregular. Despite that, it turns out that the transition is still
present, as it can be seen from the angular distributions of Fig.~\ref{fig:fig10}, corresponding to Fig.~\ref{fig:fig9}e, i.e. at a matter density
$10^5$ g/cm$^{3}$. The transition is much smoother and it is not clear if a real multi-angle regime is reached.  For the inverted hierarchy the
evolution is more regular. To be noticed that at a density slightly smaller than the critical one, where the signal disappears, the oscillations are
quite regular.

\par

\begin{figure}[t]
\begin{center}
\includegraphics[width=0.9\textwidth]{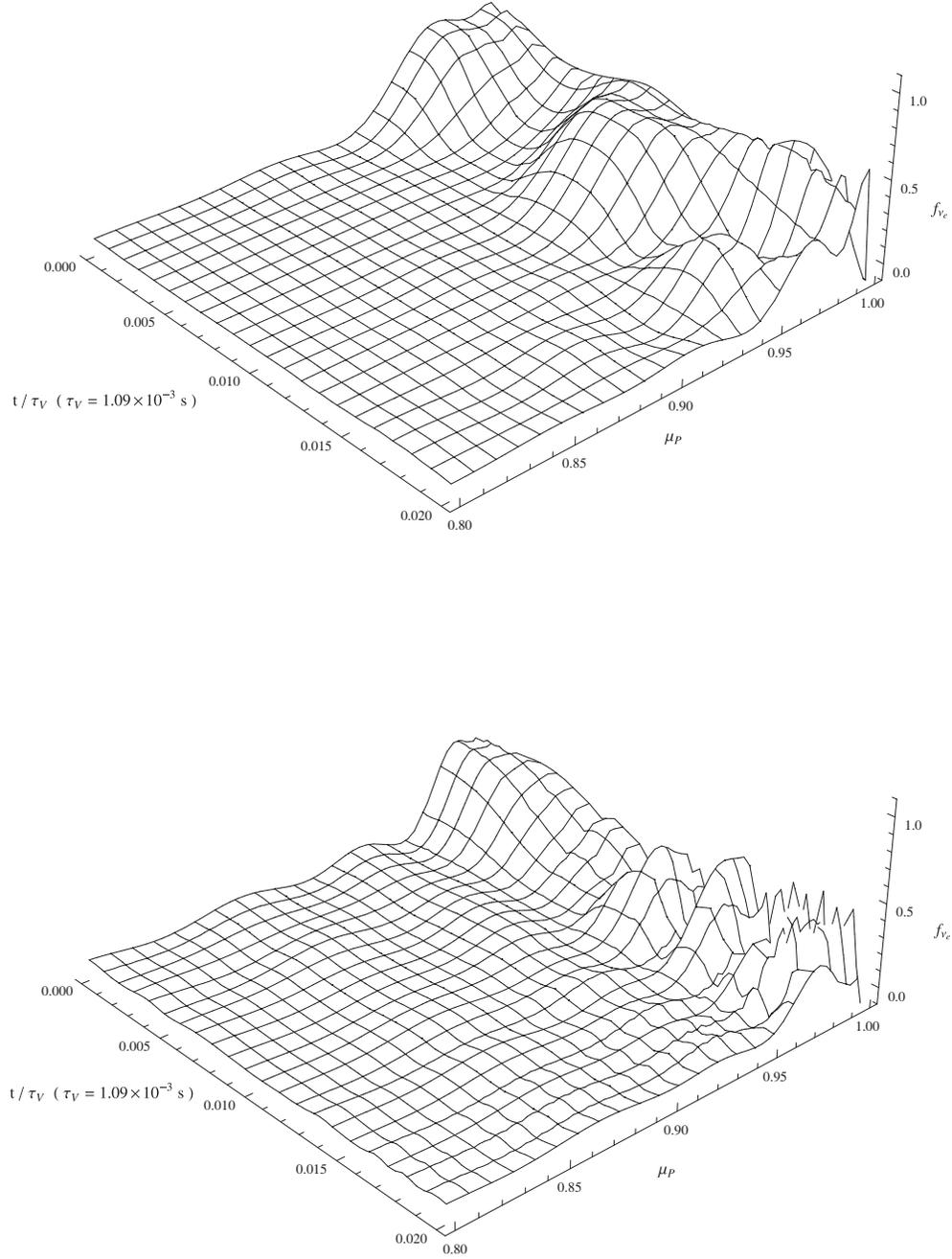}
\caption{Angular distribution of $\nu_e$ content as a function of time for a free neutrino gas with only  self-interaction at the distance of $30$ km
(upper plot) and $40$ km (lower plot). Neutrino density is fixed to $10^{32}$ cm$^{-3}$ and hierarchy is normal.} \label{fig:fig11}
\end{center}
\end{figure}

It can be interesting to look at the time evolution of the angular distribution in the case where no matter is included, i.e. for a gas of free
neutrinos with self-interaction only. The time evolution of the angular distributions corresponding to $30$ and $40$ km. are reported in
Fig.~\ref{fig:fig11}. In this case the only effect is coming from the focalization due to the distance. One can observe that there is still a
transition of regime, but it is much smoother in time than in the previous cases. The angular distribution changes smoothly from a uniform one to a
structured one. The latter, however, does not appear so random as in the previous cases but it displays only some modulations. The mixing angle has
the vacuum value, much larger than the one in the matter with electrons, and this can be the reason of this different behavior.

\par

\section{Discussion and conclusion}

For a given flavor composition and symmetric neutrino-antineutrino flavor content, we have estimated the critical matter density below which the
transition from single-angle to multi-angle behavior can be expected and the time scale of the transition for a wide set of density values. The
method is based on the expansion of the Boltzmann equations in Legendre polynomials. The calculations were performed both near the neutrino sphere
and away from the neutrino sphere. The focusing of the neutrino flow at different distances modifies both the critical density and the time scale.
Once a density profile is given, one could get an estimate of the region where the transition should first occur. Since the density profile changes
with neutrino emission time, this region will move with time. In general the matter density profile is uniformly decreasing along the emission time,
therefore the region for the transition will move toward the neutrino sphere. Furthermore since the neutrino flavor modulations are more apparent
near the neutrino sphere, the neutrino signal should be stronger at the later stage of the neutrino emission epoch. The transition appears to be
smoother as the distance increases, due to the focalization of the neutrino beam. If the neutrino flow is asymmetric, the flavor evolution changes
drastically as the asymmetry reaches a critical value, that is estimated to be not larger than $0.1$. This is in agreement with the findings in
Ref.~\cite{este}. This type of analysis is usually performed by introducing the so called ``polarization vector'', whose three-dimensional evolution
in time (or distance) gives an overall view of the flavor dynamical evolution. We preferred to use directly the full angular distribution of the
flavor content since it gives a more detailed description of the flavor structure. Of course the two methods are equivalent and correspond to two
different representations.

\par

The analysis was done within the approximation of local uniform matter for the Boltzmann equation, where the time is the only evolution parameter
\cite{raf2,raf3}. If a transition is present, it occurs in a very short time, which justifies the approximation. However another approximation method
is the stationary assumption, where the only evolution parameter is the distance from the neutrino sphere \cite{duan,stra,este,raf1}. This is
justified since the time evolution of the density profile is much longer than the neutrino transit time through the envelope. This alternative
analysis within the polynomial expansion, and its comparison to the local approximation, is left to a future work.

\section*{Acknowledgements}

\addcontentsline{toc}{sectio}{Introduction} We thank Dr. G. F. Burgio  for providing us the density profiles reported in Fig.~\ref{fig:fig1}.

\appendix*

\section{Transport with all main neutrino interactions, symmetry conditions and equations on the set of Legendre polynomials }

Recently it was developed a formalism to describe transport processes in terms of Wigner functions, which is able to incorporate the quantum
phenomenon of flavor conversion of neutrinos \cite{wign}. In terms of the matrix of equation~(\ref{F-matrix}) the complete evolution equation is

\begin{equation}\label{transport-eq}
\frac{{\partial {\cal F}}}{{\partial t}} + \frac{1}{2}\left\{
{{\bf{v}},\frac{{\partial {\cal F}}}{{\partial {\bf{r}}}}} \right\}
+ \frac{1}{2}\left\{ {{\bf{\dot p}},\frac{{\partial {\cal
F}}}{{\partial {\bf{p}}}}} \right\} =  - i\left[ {\Omega ,{\cal F}}
\right] + {\cal C}.
\end{equation}

In this equation diagonal terms ere real valued quantities that have the meaning of phase space densities, while off-diagonal terms are complex
quantities with the  meaning of macroscopic overlap functions. $\Omega$ is the mixing Hamiltonian and ${\cal C}$ is the collision matrix. In general
the Hamiltonian includes three different terms corresponding to vacuum mixing, that is energy dependent, to matter interaction, that depends on the
density profile, and the neutrino-neutrino self-interaction. The last contribution is more difficult to treat because it depends on the angle among
neutrinos impulses. Introducing real and imaginary part of off diagonal macroscopic overlap as $f_r  $ and $f_i $, one can easily find the
generalized Boltzmann equations for two neutrino flavor interacting with a background and with neutrino-neutrino interactions included, and similarly
for their antiparticles.

\par

For application it is useful rewrite transport equations in terms of
specific intensities of neutrino radiation field using the relation
connecting invariant density $f_{\nu}$ to corresponding specific
intensity $I_{\nu }$ according to $I_\nu   = E ^3 f_\nu/\left( {2\pi
\hbar } \right)^3 c^2$.

The main terms to be discussed in detail for our purposes are certainly collision term ${\cal C}_{\nu }$ and neutrino-neutrino interaction term
$B_\nu$.

Generic collision term ${\cal C}_{\nu }$ is usually written in terms of specific intensity. For convenience we report here the explicit expression
taken from Ref.~\cite{stra,bur2}

\begin{equation}
{\cal C}'_{\nu  }  =  - \kappa _{\nu  }^s I_{\nu  }  + \kappa _{\nu
 }^a \left( {\frac{{B_{\nu  }  - I_{\nu  } }}{{1 - {\cal
F}_{\nu  }^{eq} }}} \right) + \frac{{\kappa _{\nu  }^s }}{{4\pi
}}\int {\Phi _{\nu  } \left( {{\bf{\Omega }},{\bf{\Omega '}}}
\right)} I_{\nu } \left( {{\bf{\Omega '}}} \right)d\Omega ',
\end{equation}

\noindent where ${\cal C}_{\nu  }   =  \left( {2\pi \hbar } \right)^3 {\cal C}'_\nu c/E^3$.

The quantity $ \Phi _{\nu } $ is a phase function for scattering integrated over the solid angle and gives the probability that a particle enters
into the beam or go out. For a scattering process $i$ it is well approximated by

\begin{equation}
\Phi _i \left( {\Omega ,\Omega '} \right) = \Phi _i \left( {\Omega \cdot \Omega '} \right) = 1 + \delta _i \Omega  \cdot \Omega ' = 1 + \delta _i \mu
_\nu ,
\label{eq:delta}
\end{equation}

\noindent where $\delta _i$ is a constant specific to each scattering process and $ \mu_\nu $ is the angle between the incident and outgoing
neutrinos.The quantity $ \kappa _{\nu  }^a $ is the sum of all absorption processes, $n_i$ is the number density of matter species $i$ and $\sigma
_i^a$ denotes the absorption cross sections. Similarly for scattering processes one introduces the corresponding quantities $ \kappa _{\nu  }^s $ and
$\sigma _i^s$. The quantity ${\cal F}_{\nu}^{eq} $ is the equilibrium Fermi-Dirac occupation probability for neutrino species and $ B_{\nu } $ is the
corresponding black body specific intensity \cite{bur2}.

Generic neutrino self-interaction term $B_\nu$ has an integral form and using the distribution $f_\nu$ it can be written as

\begin{equation}
B_{\nu  } \left( {{\bf{p}}{\bf{,r}},t} \right) = \beta \int {d^3
{\bf{q}}} \left( {1 - \cos \theta ^{{\bf{pq}}} } \right)f_{\nu }
\left( {{\bf{q}}{\bf{,r}},t} \right),
\end{equation}

\noindent where $ \theta^ {\bf{pq}}$ is angle between a test neutrino with momentum $\bf{p}$ and a fixed background neutrino with momentum $\bf{q}$.
In this coupling coefficient integration goes over all momenta in the ensemble of neutrino background and the strength is given by $ \beta = \sqrt 2
G_F /\hbar $.

Purely kinematic aspects about neutrinos transport are in the partial derivatives at first member of the discussed equations while at second member
the dynamical terms governing flavor conversion and collisions are present. In six-dimensional phase space the single particle Boltzmann equation can
be generally expressed in compact form as the total derivative for an arbitrary $f$ equal to collision integral.

We consider the simplest case in which neutrinos are distributed
uniformly in space and can be scattered by the nucleons when
simultaneously interact with themselves and electrons. We also
restrict neutrino energy spectrum to be monochromatic.

It should be noted that restricting the energy to a single value
implies a choice for the structure of the emission term. Indeed,
this term depends on the neutrino energy and the temperature of a
specific flavor, but is not proportional to the distribution
function. Assuming the $f$ proportional to a Dirac delta function in
energy to select a fixed value in the spectrum a choice can be made
for the emission term and two different types of approximation are
possible. The first implies an integrated emission term not energy
dependent, while the latter implies energy dependence via cross
section. In this paper we fixed temperature to typical values and
use the second choice.

With all discussed approximation the time derivative of the
distributions $f$ is only term of the left-hand sides of equations
not vanishing in a fixed cartesian system of coordinates.

The mathematical structure of transport equations and the angular dependence of the densities provides an useful expansion on the set of orthogonal
Legendre polynomials. In fact different terms depend on the unknown functions with a kind of linear or quadratic dependence. Specifically the
self-interaction neutrino term $B_\nu$ has integral form and multiplies other terms containing the functions $f$, which results in quadratic
expressions. A generic $f_\nu$ for a fixed energy value has components defined as the integrals

\begin{equation}
f_\nu ^{\left( l \right)} \left( {t,E} \right) = \int_{ - 1}^1 {d\mu
_p P^{\left( l \right)} \left( {\mu _p } \right)} f_\nu  \left(
{t,E,\mu _p } \right),
\end{equation}

\noindent and it is possible to reconstruct the  original densities using the expansion

\begin{equation}
f_\nu  \left( {t,E,\mu _p } \right) = \sum\limits_{l = 0}^\infty
{\left( {l + \frac{1}{2}} \right)} P^{\left( l \right)} \left( {\mu
_p } \right)f_v^{\left( l \right)} \left( {t,E} \right).
\end{equation}

The typical time scale characterizing neutrino self-interactions is the period associated with bipolar oscillations, $ \tau _B = 2\pi /\sqrt {2\omega
\mu } $, where $\mu=\beta n_\nu$ with $n_\nu$ the numerical density of all neutrinos species. It is also useful to introduce the period of free
oscillation in vacuum $ \tau _V  = L/c $ in terms of oscillation length $ L = 4\pi \hbar E/\Delta m^2 c^3 $ depending on energy and neutrino mass
square difference. For convenience we choose $\tau_V$ as unit of time, and define a specific interaction rate due to collisions in terms of times $
\tau _i^s = 1/cN_A \rho Y_i\sigma _i^s $ for scattering and $ \tau _i^a = \left( {1 - {\cal F}_{\nu}^{eq} } \right)/cN_A \rho Y_i \sigma _i^a$ for
stimulated absorption-emission. The quantity $Y_i$ are fractions of nucleons, $ \sigma _i^s $ and $ \sigma _s^a $ are cross sections. Index $i$
indicates nucleon and can be equal to $n$ for neutrons, $p$ for protons and $t$ for the sum of processes, with  $ {\tau _t^s } ^{ - 1}= {\tau _p^s }
^{ - 1}+ {\tau _n^s } ^{ - 1} $. Note that interaction rates are not flavor independent because the cross sections depend on the mass of the lepton
specifically involved in the process.

With above assumptions, if all distribution and overlap functions are normalized to the neutrinos density, the equations can be rewritten in a form
useful for their numerical solution. For neutrinos and antineutrinos one gets a set of eight coupled nonlinear equations. For example, the first of
these has the explicit form

\begin{equation}
\label{eq1_legendre}
\begin{array}{*{20}c}
   {\dot f_{\nu _e }^{\left( l \right)}  = 2\pi f_i^{\left( l \right)} \sin 2\theta  + }  \\
   {}  \\
   { + 2\pi \left( {\frac{{\tau _V }}{{\tau _B }}} \right)^2 \left( {\left( {f_r^{\left( 0 \right)}  - \tilde f_r^{\left( 0 \right)} } \right)f_i^{\left( l \right)}  - } \right.\left( {f_i^{\left( 0 \right)}  - \tilde f_i^{\left( 0 \right)} } \right)f_r^{\left( l \right)}  + }  \\
   {}  \\
   { - a_l \left( {f_r^{\left( 1 \right)}  - \tilde f_r^{\left( 1 \right)} } \right)f_i^{\left( {l + 1} \right)}  + b_l \left( {f_i^{\left( 1 \right)}  - \tilde f_i^{\left( 1 \right)} } \right)f_r^{\left( {l - 1} \right)}  + }  \\
   {}  \\
   { - b_l \left( {f_r^{\left( 1 \right)}  - \tilde f_r^{\left( 1 \right)} } \right)f_i^{\left( {l - 1} \right)}  + a_l \left( {f_i^{\left( 1 \right)}  - \tilde f_i^{\left( 1 \right)} } \right)\left. {f_r^{\left( {l + 1} \right)} } \right) + }  \\
   {}  \\
   { - \left( {\frac{{\tau _V }}{{\tau _t^s }}} \right)f_{\nu _e }^{\left( l \right)}  + \left( {\frac{{\tau _V }}{{\tau _n^a }}} \right)\left( {B^ *  \left( {T_{\nu _e } } \right)\delta _{l,0}  - f_{\nu _e }^{\left( l \right)} } \right) + }  \\
   {}  \\
   { + \left( {\frac{{\tau _V }}{{\tau _p^s }}} \right)\left( {f_{\nu _e }^{\left( 0 \right)} \delta _{l,0}  + \frac{{\delta _p }}{3}f_{\nu _e }^{\left( 0 \right)} \delta _{l,1} } \right) + \left( {\frac{{\tau _V }}{{\tau _n^s }}} \right)\left( {f_{\nu _e }^{\left( 0 \right)} \delta _{l,0}  + \frac{{\delta _n }}{3}f_{\nu _e }^{\left( 0 \right)} \delta _{l,1} } \right)}.  \\
\end{array}
\end{equation}

\noindent The angle $ \theta $ is the vacuum mixing angle between two neutrino species and the density of neutrinos appears due to the normalization.
The quantity ${B^*  \left( {T_{\nu _e } } \right)}$ is the integrated emission term depending on temperature according to the above discussion. The
choice of time unit is convenient since the ratio $\tau _B /\tau _V $ is dependent only on the ratio $ \omega /\mu $.

In four of the eight equations a parameter $A$ appears, that represents the potential induced by ordinary matter. It is proportional to concentration
of charge lepton according to $ A = 2\sqrt 2 LG_F n_e/\left( {\pi \hbar c} \right)$.  Note that above equations have some terms with different
algebraic signs from the original reference. This depends on different convention used for the relative sign between the real and imaginary parts of
complex functions associated with neutrinos and antineutrinos.

\vfill\eject

\centering\tabcolsep=10mm
\begin{table}
\begin{tabular}{cc}
\hline {\em Parameter} & {\em Value}\\ \hline
$\Delta {m^2}$ & $7.59\cdot10^{ - 5}\rm{~eV^2}$ \\
$\sin \left( {2\theta } \right)$ & $0.93$ \\
$T_{\nu _e}$ & 2.76{\rm{ MeV}} \\
$T_{\nu _\mu }$ & 6.26{\rm{ MeV}} \\
$T_{ \overline{\nu} _e}$ & 4.01{\rm{ MeV}} \\
$T_{ \overline{\nu} _\mu }$ & 6.26{\rm{ MeV}} \\
$\delta_p$ & -0.2 \\
$\delta_n$ & -0.1 \\
\hline
\end{tabular}
\caption{\label{tb:tab}Neutrino physical parameters used in the simulations. For the meaning of the parameters $\delta$ see Eq.~(\ref{eq:delta}). }
\end{table}

\clearpage

\vfill\eject

\vfill\eject

\vfill\eject

\vfill\eject

\vfill\eject

\vfill\eject

\vfill\eject

\vfill\eject

\vfill\eject

\vfill\eject

\vfill\eject

\end{document}